\documentclass[prl,twocolumn,superscriptaddress,showpacs]{revtex4}
\usepackage{amssymb,amsmath}
\usepackage{latexsym}
\usepackage{graphicx}
\usepackage{hyperref}
\def\sro{SrRuO$_3$}
\def\ngo{NdGaO$_3$}
\def\sto{SrTiO$_3$}

\begin{document}

\title{Fractional power-law conductivity in \sro\ and its consequences}

\author{J.~S.~Dodge}
\author{C.~P.~Weber}
\author{J.~Corson}
\author{J.~Orenstein}
\affiliation{Department of Physics, University of California at
  Berkeley, Berkeley, California 94720}
\affiliation{Materials Sciences Division, E.~O.~Lawrence Berkeley
  National Laboratory, Berkeley, California 94720}

\author{Z.~Schlesinger}
\affiliation{Department of Physics, University of California at Santa
  Cruz, Santa Cruz, California 95064}

\author{J.~W.~Reiner}
\author{M.~R.~Beasley} 
\affiliation{Edward L.~Ginzton Laboratories, Stanford University,
  Stanford, California 94305}

\date{\today}

\begin{abstract}
  We combine the results of terahertz time-domain spectroscopy with
  far-infrared transmission and reflectivity to obtain the
  conductivity of \sro\ over an unprecedented continuous range in
  frequency, allowing us to characterize the approach to zero
  frequency as a function of temperature.  We show that the
  conductivity follows a simple phenomenological form, with an
  analytic structure fundamentally different from that predicted by
  the standard theory of metals.
\end{abstract}

\pacs{72.15.Qm,78.66.Bz,75.20.En,74.20.Mn}
\maketitle

One of the most exciting proposals to emerge from the study of
high-$T_{c}$ superconductors is that Landau's Fermi liquid theory
(FLT) breaks down in the metallic state above
$T_{c}$~\cite{Anderson1987}.  This would have profound implications,
since FLT provides the foundation for our current understanding of
metals, together with systems as diverse as liquid $^3$He and nuclear
matter~\cite{PineNoz}.  Evidence for its breakdown in high-$T_{c}$
superconductors comes from a variety of experiments, including
photoemission, electrical transport and optics~\cite{Orenstein2000}. 
More recently similar evidence has been found in other
compounds~\cite{Imada1998}.  Here we show that the complex optical
conductivity $\sigma(\omega,T)$ of one such material, the
ferromagnetic metal
\sro~\cite{Bozovic1994,Klein1996l,Klein1996r,Kostic1998}, behaves
according to a remarkably simple power-law form, which deviates
sharply from the prediction of FLT. This observation provides valuable
insight into the nature of charge scattering in unconventional metals.

According to FLT, the qualitative properties of an interacting
electron gas are the same as those of a noninteracting gas, if probed
on a sufficiently low-energy scale.  The optical conductivity of a
system of noninteracting, charge $e$ carriers obeys the Drude form,
$\sigma(\omega)=(ne^2/m)/(1/\tau-i\omega)$, where $n$ is the carrier
density and $m$ and $\tau$ are the effective mass and scattering time
of the carriers.  FLT predicts that $\sigma$ at low $\omega$ remains
of the Drude form in the presence of interactions, with a spectral
weight which decreases as interactions increase.  The {\it f\,}-sum
rule dictates that the total spectral weight is conserved, so that
spectral weight must shift to higher energies.  This additional
component to $\sigma(\omega)$ is known as the incoherent part of the
intraband conductivity.

Infrared reflectivity studies indicate that both high-$T_{c}$
superconductors and \sro\ exhibit conductivity with an anomalous
power-law dependence on frequency, $\sigma_{1}(\omega) \propto
\omega^{-\alpha}$, with $\alpha \sim 0.5$ for \sro~\cite{Kostic1998}
and $\alpha\sim 0.7$ in the high-$T_c$
materials~\cite{Schlesinger1990,ElAzrak1994}.  The Drude form yields
$\sigma_{1}\propto\omega^{-2}$ at comparable frequencies.  If FLT is
valid, the conductivity in excess of Drude must be identified with
interband transitions or the incoherent component of the spectrum, and
there must be a crossover at lower frequency to the renormalized Drude
conductivity~\cite{Tanner1992,Georges1996}.

Recently, Ioffe and Millis suggested that the entire conductivity
spectrum of the high-$T_{c}$ materials could be understood as a single
component, rather than two, as in FLT~\cite{Ioffe1998}.  The spectrum
they derived was generalized by van der Marel to the following useful
form~\cite{vanderMarel1999},
\begin{equation}
  \label{eq:powerlawsig}
  \sigma(\omega) = \frac{A}{(1/\tau - i\omega)^{\alpha}},
\end{equation}
where $A$ is a parameter with units that depend on the value of
$\alpha$. Ioffe and Millis obtained $\alpha = 1/2$ by assuming that
the carrier lifetime depends strongly on its direction of
motion~\cite{Carrington1992,Hlubina1995,Stojkovic1996,Xiang2000}.  In
the limit $1/\tau \rightarrow 0$, Eq.~\ref{eq:powerlawsig} is similar
to one derived by Anderson~\cite{Anderson1997}, under different
assumptions.  Note that Eq.~\ref{eq:powerlawsig} includes the Drude
form as a special case, with $\alpha = 1$.  In allowing $\alpha$ to
deviate from unity, we obtain the observed power law dependence of the
conductivity on frequency.  However, we also subvert a standard
assumption of transport physics, that the conductivity at $\omega = 0$
is proportional to a scattering time. Instead, the dc conductivity is
proportional to a fractional power of a scattering time, that is,
$\sigma_{dc} = A\,\tau^\alpha$.  Moreover, the analytical structure of
$\sigma(\omega)$ changes, from having a simple pole at $\omega =
-i/\tau$, to being multiple-valued with a branch point there.

Although the models described above are distinct, they may be
difficult to distinguish experimentally if the Drude component
predicted by FLT is masked by the incoherent conductivity.  The most
stringent test of FLT is at low temperature, where the Drude
component, if one exists, would be sharpest.  In the cuprates, the
temperature range over which the normal state conductivity can be
studied is limited by the onset of superconductivity.  The absence of
superconductivity in \sro\ permits a much more meaningful test of FLT
than previously available.

We find that $\sigma(\omega,T)$ of \sro\ at low temperature is
described well by Eq.~\ref{eq:powerlawsig} over nearly three decades
in $\omega$ (6--2400~cm$^{-1}$) with $\alpha \sim 0.4$, in strong
disagreement with FLT.  Our results at frequencies below 100~cm$^{-1}$
were obtained from transmission measurements on thin film samples of
\sro\ grown epitaxially on \ngo\ substrates~\cite{AhnPhD,Eom1992}.  To
probe the region where $\omega \sim 1/\tau$, we used conventional
Fourier-transform infrared spectroscopy to measure the transmittance
($\mathcal{T}$).  For the range $\omega \ll 1/\tau$, we used
time-domain terahertz spectroscopy (TDTHz) to measure the complex
transmission amplitude $t(\omega,T)$ in the millimeter wave region of
the spectrum.  The residual resistivity in these films is typically
50~$\mu\Omega-$cm.  \sto\ substrates produce \sro\ films with lower
resistance, but this substrate material has such a large
temperature-dependent dielectric constant that an accurate
determination of the conductivity from transmission measurements is
prohibitively difficult.  At frequencies above 100~cm$^{-1}$, we
measure the reflectivity from a thick film of \sro\ deposited on a
\sto\ substrate~\cite{Kostic1998}.  We have derived $\sigma(\omega,T)$
from each of these measurements, as described below.

\begin{figure}[tbp]
  \includegraphics[width=.923\columnwidth]{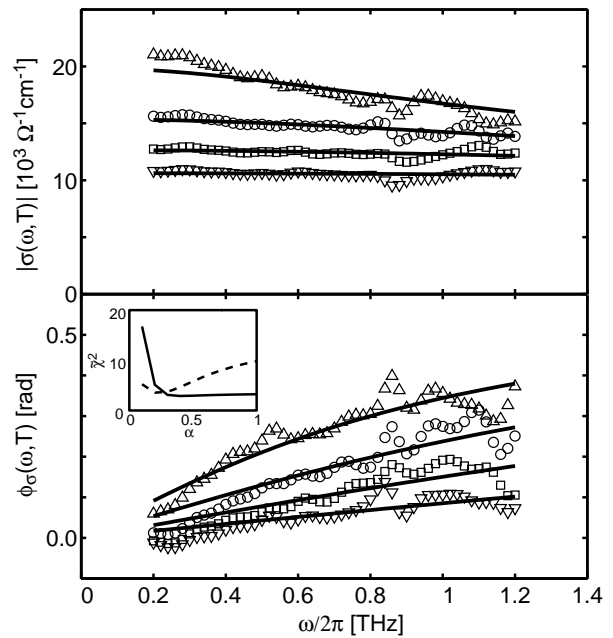}
    \caption{Measured amplitude (upper panel) and phase (lower panel) of 
      $\sigma(\omega,T)$ in \sro\ at four representative temperatures.
      $\bigtriangleup$: T = 8~K, $\bigcirc$: 40~K, $\Box$: 60~K, and
      $\bigtriangledown$: 80~K. Lines are fits to the data using
      Eq.~\ref{eq:powerlawsig}, with $\alpha = 0.4$.  The inset shows
      the reduced $\chi^{2}$ error associated with the phase (solid
      line) and amplitude (dashed line) fits to
      Eq.~\ref{eq:powerlawsig}, as a function of $\alpha$.}
\label{fig:swfits}
\end{figure}

We use TDTHz to measure the complex transmission amplitude $t(\omega)$
of \sro\ in the range 0.2--1.2 THz (6--36 cm$^{-1}$)~\cite{Nuss1998}. 
We compared the \sro\ film on its substrate to a bare \ngo\ substrate
at each temperature, using a vapor flow cryostat with a translating
sample mount.  The ratio of the two complex transmission amplitudes,
$t_{sr}(\omega) = t_{sample}(\omega)/t_{ref}(\omega)$, is a simple
function of the substrate index $n$, $\sigma(\omega)$ of the film, and
the film thickness $d$:
\begin{equation}
  \label{eq:transamp}
  t_{sr}(\omega) = \frac{n+1}{n + 1 + \sigma(\omega) Z_0 d}.
\end{equation}
$Z_0$ is the impedance of free space.  We inverted
Eq.~\ref{eq:transamp} to obtain both the amplitude and phase of the
complex conductivity as a function of temperature and frequency, which
we show in Fig.~\ref{fig:swfits} for our most thoroughly studied \sro\
film.
Also shown are the results of the best fit to Eq.~\ref{eq:powerlawsig}
with $\alpha = 0.4$, following the procedure described below.  The
statistical error on a typical measurement of the conductivity phase
$\phi_{\sigma}(\omega,T) = \arg[\sigma(\omega,T)]$ is $\pm 0.02$~rad
at 1~THz, two orders of magnitude better than that obtained in typical
reflectivity measurements at this frequency.  The temperature
dependence of the effective path length in the substrate and cryostat
windows provides the largest source of systematic error, which we
estimate to be less than $\pm 0.03$~rad at 1~THz.  Errors in the
measurement of the conductivity amplitude are dominated by statistical
uncertainty in the transmission amplitude, which is typically $\pm
3$\%.

To obtain an objective best fit of Eq.~\ref{eq:powerlawsig} to the
TDTHz measurements, we first perform a least-squares fit of the
measured conductivity phase to $\phi(\omega) =
\alpha\tan^{-1}(\omega\tau)$ at each temperature, for several values
of $\alpha$.  This allows us to determine $\tau(T;\alpha)$, the best
fit value of $\tau$ for each temperature, with different assumed
values of $\alpha$.  Unlike a global fit to $\sigma(\omega)$, this
procedure is independent of the conductivity amplitude.  Next we
performed a separate least-squares fit of the amplitude to
$|\sigma(\omega)| = A/(1/\tau^2 + \omega^2)^{\alpha/2}$, using the
$\tau = \tau(T;\alpha)$ from the phase fits and allowing only $A$ to
vary.  The quality of these fits~\cite{ReducedChi} are shown as a
function of $\alpha$ in the inset to Fig.~\ref{fig:swfits}.  The phase
fits exhibit a weak optimum at $\alpha = 0.4$, with a sharp decrease
in quality below $\alpha = 0.3$ but with relatively little change in
quality as $\alpha$ increases from 0.4 to unity.  The amplitude fits,
on the other hand, worsen dramatically as $\alpha$ changes from 0.3 to
unity.  Taken together, these fits allow us to limit the range of
acceptable values for $\alpha$ to 0.2--0.5.

\begin{figure}[tbp]
  \includegraphics[width=.923\columnwidth]{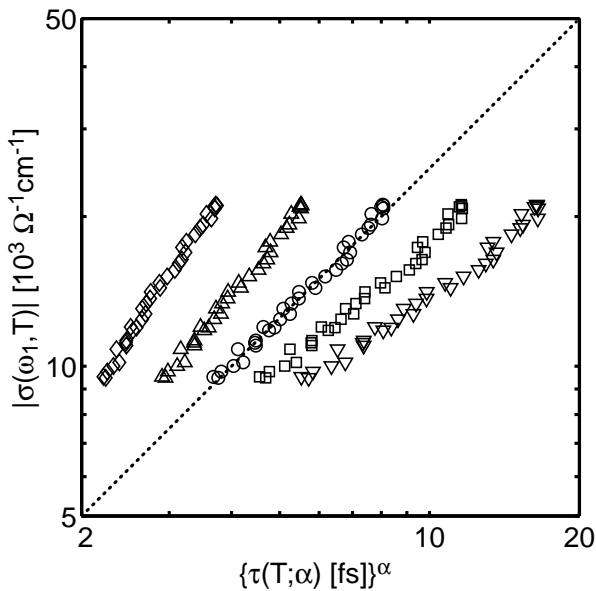}
    \caption{Logarithmic plot demonstrating the scaling relationship of
      $\sigma(\omega_{1},T)$, $\omega_{1}/2\pi = 0.2$~THz, and
      $\tau(T;\alpha)$ obtained by fitting to
      Eq.~\ref{eq:powerlawsig}.  The plot is parametric in temperature
      from 5--92.5~K in steps of 2.5~K, for various choices of
      $\alpha$: $\Diamond$, 0.2; $\bigtriangleup$, 0.3; ,$\bigcirc$,
      0.4; $\Box$, 0.5; and $\bigtriangledown$, 0.6.  The dotted line
      is given by $\sigma(\omega=0) \propto \tau^\alpha$.}
\label{fig:sigvstau}
\end{figure}

We develop this analysis further in Fig.~\ref{fig:sigvstau}, which
shows a logarithmic plot of the conductivity amplitude at our lowest
frequency, $\omega_{1}/2\pi = 0.2$~THz, versus
$[\tau(T;\alpha)]^{\alpha}$, with $\tau$ in femtoseconds.
The plot is parametric in temperature for several different values of
$\alpha$.  In Eq.~\ref{eq:powerlawsig}, $\sigma(\omega\rightarrow 0)
= A\, \tau^{\alpha}$; if $A$ remains constant with temperature,
Eq.~\ref{eq:powerlawsig} will yield a straight line with unity slope
on this plot, indicated by the dotted line.  The best fit to this
slope is obtained for $\alpha = 0.4$, and already at $\alpha = 0.6$ a
clear deviation is observed.  At higher values of $\alpha$ the slope
decreases yet further, so the case $\alpha = 1$ corresponding to Drude
conductivity requires $A$ to increase strongly with temperature.
Thus, $\alpha = 0.4\pm.1$ provides not only the best fit to
Eq.~\ref{eq:powerlawsig}, but also the most compact description of the
data.

\begin{figure}[tbp]
  \includegraphics[width=.923\columnwidth]{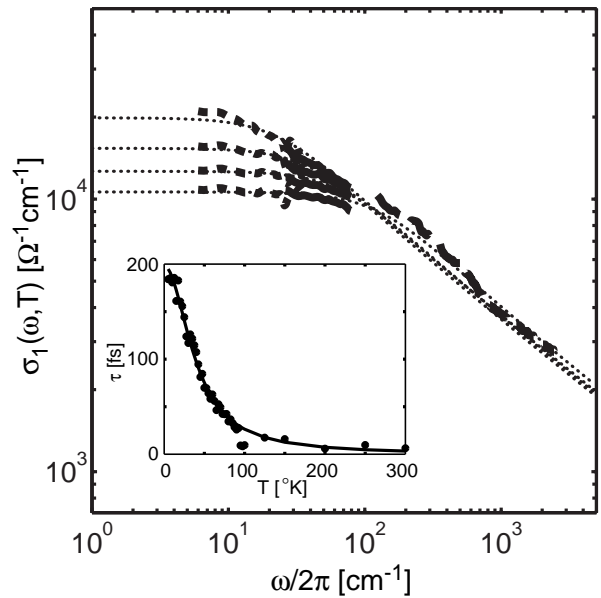}
    \caption{Logarithmic plot of the conductivity obtained by three
      methods, in three ranges of frequency.  The conductivity
      obtained from the infrared reflectivity at 40~K is indicated by
      the long dashed line.  Results from far-infrared transmission
      measurements, as described in the text, are indicated by solid
      lines, and TDTHz measurements by short dashed lines, with
      both sets ordered in temperature from top to bottom with $T =
      8$~K, 40~K, 60~K, and 80~K. Least-squared fits to
      Eq.~\ref{eq:powerlawsig} using only TDTHz data are shown by
      dashed lines.  Inset: temperature dependence of $\tau$ obtained
      for $\alpha = 0.4$ (closed circles), compared to
      Eq.~\ref{eq:quadtemp} (solid line).}
\label{fig:allsigcomp}
\end{figure}

In Fig.~\ref{fig:allsigcomp} we show $\sigma_{1}(\omega)$ of \sro\ 
over two and a half decades in frequency, obtained in three separate
measurements.
Data in the lowest frequency range is taken from TDTHz measurements at
the same four representative temperatures shown in
Fig.~\ref{fig:swfits}, and extends from 6--36~cm$^{-1}$.  In the
intermediate frequency range 26--80~cm$^{-1}$, we have measured
${\mathcal T}(\omega,T)$, which is a real quantity and therefore
incapable of providing the complex conductivity without further
analysis.  Using the $\tau(T)$ obtained from TDTHz for $\alpha = 0.4$,
we have calculated the conductivity phase expected in this frequency
range at each temperature, then used these phase values to calculate
the conductivity amplitude directly from ${\mathcal T}(\omega,T)$.
The results of this procedure are shown for the same four temperatures
as the TDTHz data.  The continuity of the results at the crossover
frequency of these two distinct measurements may be taken as an
indication of the high accuracy with which we have determined the
conductivity.  At frequencies probed by infrared reflectivity, the
conductivity is relatively temperature independent below 100~K, and we
show only one measurement taken at 40~K~\cite{Kostic1998}.

With the parameters obtained from the TDTHz data,
Eq.~\ref{eq:powerlawsig} may be used to predict the behavior of
the conductivity at all frequencies and temperatures.  The dotted
lines in Fig.~\ref{fig:allsigcomp} show the conductivity calculated
from Eq.~\ref{eq:powerlawsig}, using the parameters obtained from
TDTHz.  With a single global parameter $A$ and a single
temperature-dependent parameter $\tau(T)$, this model fits the data
exceptionally well, at frequencies two orders of magnitude higher than
those at which the parameters were obtained.  As we increase the
temperature above 95~K, both our measurements and earlier reflectivity
measurements begin to deviate from the form discussed here, and
develop a pseudogap structure which may be related to the transition
from ferromagnetism to paramagnetism~\cite{Kostic1998}.  We leave a
detailed discussion of this behavior to a later publication, limiting
our discussion here to temperatures below 95~K, deep within the
ferromagnetic state.

As shown in Fig.~\ref{fig:allsigcomp}, $1/\tau$ sets the frequency
scale at which the $\omega^{-\alpha}$ divergence is cut off, forcing
$\sigma_{1} \propto \tau^{\alpha}$ in the dc limit.  Thus the
ubiquitous practice of inferring scattering times from electrical
transport via $\sigma_{dc} \propto \tau$ is erroneous, whenever the
conductivity behaves as Eq.~\ref{eq:powerlawsig} with $\alpha \neq 1$. 
The resistivity of \sro\ for 25~K~$\lesssim T \lesssim$~120~K exhibits
approximately linear temperature dependence, which then crosses over
at lower temperatures to become constant as impurity scattering
dominates~\cite{Klein1996r}.  In our analysis, this implies a
scattering rate with at least a quadratic dependence on
temperature.

A simple form which approximates the observed behavior is
\begin{equation}
    \frac{\hbar}{\tau(T)} = \frac{\hbar}{\tau_{0}} +
    \frac{k_{B}T^{2}}{T_{0}},
    \label{eq:quadtemp}
\end{equation}
with a temperature independent, or elastic, scattering time $\tau_{0}$
and a characteristic temperature $T_{0}$~\cite{Ioffe1998}.  Here,
$\hbar$ is Planck's constant, and $k_{B}$ Boltzmann's constant. This
gives a temperature dependent resistivity, $\rho(T) = A + BT^2$,
observed in the highest quality films at low
temperatures~\cite{Mackenzie1998}. The inset to
Fig.~\ref{fig:allsigcomp} shows $\tau(T;0.4)$ together with the best
fit to Eq.~\ref{eq:quadtemp}, with $\tau_{0} = 198$~fs and $T_{0} =
40$~K.  The agreement is quite good over the entire temperature range,
although in the region $T > 95$~K $\tau$ is comparable to our
measurement accuracy.

It is interesting to note that despite the discrepancy of
Eq.~\ref{eq:powerlawsig} with FLT, the functional form of
Eq.~\ref{eq:quadtemp} is exactly what FLT would predict for $\tau$,
though with $T_0$ at least two orders of magnitude larger than 40 K.
The origin of this temperature dependence, and of the low energy scale
$T_0$ which controls it, remains one of the important open questions
raised by these measurements.  Similar behavior has recently been
observed in the normal state of
Bi$_{2}$Sr$_{2}$CaCu$_{2}$O$_{8}$~\cite{Corson2000}, and we expect
that a considerable body of transport work will require reanalysis in
light of the observations presented here.  For example, Shubnikhov-de
Haas oscillations have been observed in \sro, with amplitudes that
display the temperature dependence of a Fermi
liquid~\cite{Mackenzie1998}.  It would be interesting to develop
theories which produce Eq.~\ref{eq:powerlawsig}, to address the
existence of these quantum oscillations.

In summary, we have studied in detail the complex conductivity of
\sro\ at low frequencies and temperatures, and shown that it agrees
well with the simple phenomenological form given in
Eq.~\ref{eq:powerlawsig}. The difference between $\alpha \sim 0.4$
observed here and the Drude form expected from FLT, with $\alpha = 1$,
is reflected in the interpretation of $\tau$, one of the fundamental
parameters in the transport theory of metals.  We have described how
this quantity influences the measured conductivity, but proper
interpretation of its microscopic meaning must await further analysis.
In particular, careful studies of the dependence of $\tau$ on
substitutional impurities or structural disorder would help to reveal
its physical origin~\cite{Ioffe1998}.

\begin{acknowledgments}

  J.~S.~D.~thanks Daniel Chemla for encouragement in pursuing this
  work.  This work was supported by the Director, Office of Energy
  Research, Office of Basic Energy Sciences, Division of Materials
  Sciences of the U.~S.~Department of Energy under Contract No. 
  DE-AC03-76SF00098, by the Stanford NSF-MRSEC Program, and by NSF
  grants DMR-0071949 and 9870252.
\end{acknowledgments}


\begin{thebibliography}{10}
\bibitem{ReducedChi}
\bibinfo{note}{The reduced $\chi^{2}$ values shown are obtained by fixing
  $\alpha$, finding the $\chi^{2}$ error on fits to Eq.~\ref{eq:powerlawsig}
  for all temperatures from 5--92.5~K and all frequencies from 0.2--1.2 THz,
  then dividing by the total number of degrees of freedom, equal to the number
  of data points minus the number of free parameters in the fit.}
\end{thebibliography}
\end{document}